\documentclass[final]{elsart}

\usepackage{epsfig}
\usepackage{psfrag}

\usepackage{amssymb}
\usepackage{amsmath}

\newcommand{\mystack}[2]{\mbox{\scriptsize$
\left(\begin{array}{c} #1 \\[-1mm] #2 \end{array}\right)$}}

\begin{document}
\hfill{\small FZJ--IKP--TH--2007--11}
\begin{frontmatter}
\title{Remarks on $NN\to NN\pi$ beyond leading order}

\author{C. Hanhart}
\ead{c.hanhart@fz-juelich.de}
and
\author{A. Wirzba}
\ead{a.wirzba@fz-juelich.de}

\address{Institut f\"{u}r Kernphysik,\\
         Forschungszentrum J\"{u}lich GmbH,\\
         D--52425 J\"{u}lich, Germany}

\begin{abstract}
In recent years a two-scale expansion was
established to study reactions of the type
$NN\to NN\pi$ within chiral perturbation theory.
Then the diagrams of some subclasses
that are invariant 
under the choice of the pion field no longer
appear at the same chiral order. In this letter
we show that the proposed expansion still leads to well
defined results.   
We also discuss the appropriate choice of
the heavy baryon propagator.
\end{abstract}


\end{frontmatter}

\section{Introduction}

\begin{figure}[t]
\begin{center}
\epsfig{file=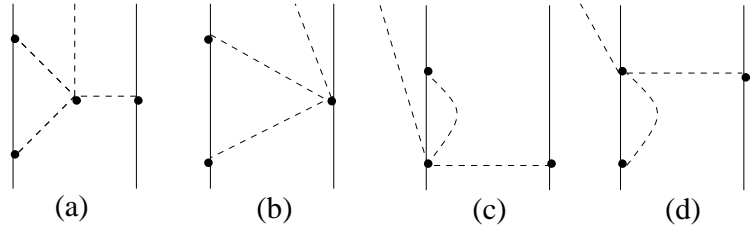,width=10cm}
\end{center}
\caption{Some one-loop diagrams that start to contribute
to $NN\to NN\pi$ at NLO, $(a)$ \& $(b)$,
and N$^4$LO, $(c)$ \& $(d)$. Dashed lines
denote pions and solid lines denote nucleons. The exchange diagrams are not
shown. Note that the diagrams $(b)$,  $(c)$ and  $(d)$ result from diagram
$(a)$ under the removal of one internal pion line.}
\label{loops}
\end{figure}

Pion production in nucleon-nucleon ($NN$) collisions is subject of theoretical
and experimental investigations already since the 1960s --- for a review of
the history of the field see Ref.~\cite{garmiz}.  However, when new high
precision data became available due to advanced accelerator technology in the
beginning of the 1990s it became clear that all phenomenological studies
performed so far were not capable of describing the data.  Several mechanisms
were proposed to cure the problem; however, no clear picture
emerged~\cite{report}.

There was the hope that chiral perturbation theory (ChPT) could resolve
the issue. As the effective field theory for the standard model
at low energies it should provide a framework to investigate
the reactions $NN\to NN\pi$ in a field-theoretically consistent
way. In a first attempt a scheme proposed by Weinberg to 
study elastic and inelastic pion reactions on nuclei~\cite{wein92}
was applied to investigate also pion production in $NN$ collisions.
 However, in doing so up to next--to--leading order (NLO) 
the discrepancy between the
calculations and data became even worse~\cite{firsts}. In addition, 
loop contributions, formally of order NNLO, 
gave even larger 
effects~\cite{loop_dmit,loop_ando} shedding doubts on an applicability of 
chiral perturbation theory to $NN\to NN\pi$.

In parallel, already in Refs.~\cite{bira} it was stressed
that the large momentum transfer, typical for meson
production in $NN$ collisions, needs to be taken care
of in the power counting. This idea was
further developed in Refs.~\cite{pwaves,mitnorbert}. 
The appropriate expansion parameter for $NN\to NN\pi$ therefore is
\begin{equation}
  \chi_{\rm prod}=p_{\rm thr}/M=\sqrt{{m_\pi}/{M}} \ ,
\label{chiprod}
\end{equation}
where $p_{\rm thr}=\sqrt{Mm_\pi}$ denotes the threshold
momentum for pion production in $NN$ collisions. $M$ and $m_\pi$ are the
masses of  the nucleon and pion, respectively.   
Here the leading-order (LO) scales as
${\mathcal O}(\chi_{\rm prod}^1)$ and subleading orders N$^n$LO scale as
${\mathcal O}(\chi_{\rm prod}^{n+1})$. For the most recent 
developments for the reaction $NN\to NN\pi$ within chiral perturbation
theory we refer to Refs.~\cite{pp2dpi}.

Thus in the reactions $NN\to NN\pi$ one is faced with a 
two-scale expansion, since both $m_\pi$ as well as $p_{\rm thr}$
appear explicitly in the expressions. For tree-level diagrams
this does not cause any problem. To perform the power counting
for loop integrals, however, a rule has to be given what scale
to assign to the components of the loop momentum. 
After subtraction of the nucleon mass $M$, the residual energy of each
external nucleon at threshold
is  $m_\pi/2$, whereas the corresponding
momentum is of order $p_{\rm thr}$. One therefore would be tempted to
take over this scaling also for the loop momentum. On the other hand
the new power counting is based on two scales, 
$p_{\rm thr}\gg m_\pi$, 
and the pions in loops are off-shell. Therefore there is no reason
why the scaling of the pion energies in loops should be different from the
scaling of the pertinent three-momenta.
In Appendix E of Ref.~\cite{report} it is shown that for all diagrams
that do not have a two-nucleon cut, each component of the 
loop momentum should be counted of order of the largest 
external momentum in the loop. 
The argument there is based on the observation that in time
ordered perturbation theory (TOPT) there is no ambiguity
for the order assignment of energies since it is a
3 dimensional theory in the first place. On the
other hand the leading order of a given TOPT amplitude
should agree to that of the corresponding Feynman amplitude.
This allows to identify the proper scale for the
energy of the loop momentum. The assignment was checked
by explicit calculations in Refs.~\cite{loop_dmit,mitnorbert}.
As a result, all components of the 
loop momentum in  diagram (a) and (b) of Fig.~\ref{loops} scale as
 $\chi_{\rm prod}M$, but those of diagram (c) and (d)
scale as $m_\pi\sim\chi_{\rm prod}^2 M$ and are therefore
suppressed\footnote{In this case
the large momentum $p_{\rm thr}$ can be removed from 
the integral by, e.g., a proper choice of the integration
variable in full analogy to the discussion given below
Eq. (\ref{firststep}).}.
One further 
consequence of the presence of two scales in the problem
is that the individual loops no longer contribute to only a single
order, but each loop contributes to infinitely many orders, since
$m_\pi/p_{\rm thr}= \chi_{\rm prod}$ appears as the argument of
non--analytic functions. The power counting only identifies the lowest
order where the particular loop starts to contribute~\cite{mitnorbert}.

In Ref.~\cite{novel} it was shown that
the sum of all diagrams of Fig.~\ref{loops}
is independent of the choice of the pion field.
However, based on the scheme developed in Refs.~\cite{pwaves,mitnorbert}
only diagram $(a)$ and $(b)$ contribute at NLO 
whereas
diagrams $(c)$ and $(d)$ start to contribute not until
order N$^4$LO 
(see Table 11 of Ref.~\cite{report}). 
The main purpose of this
letter is to investigate the consistency of these two
statements.

As we go along we also need to discuss the appropriate choice
of the nucleon propagator in the heavy baryon formulation.
This is done in Section~\ref{subleading}. Section~\ref{conclusion} contains
our conclusions.
Moreover, for clarification
two appendices are added, one
is about reparameterizations of the chiral matrix $U$, the other is about 
the $1/M$ expansion of the nucleon propagator.

\section{Dependence on the pion field to NLO}

The Lagrangian relevant for our study
may be written as~\cite{GSS}
\begin{equation}
{\mathcal L}=\frac{f_\pi^2}{4}\left<u^\mu u_\mu\right>
+\frac{f_\pi^2}{4}\left<\chi_+\right>+
\bar \Psi \left(i\gamma^\mu D_\mu -M+\frac{g_A}{2}\gamma^\mu
u_\mu\gamma_5\right)\Psi \ .
\label{LGSS}
\end{equation}
Here $\left< \dots \right>$ denotes a trace in the isospin-space,
$\Psi$ is the relativistic spinor of the nucleon, 
$D_\mu$ is its covariant
derivative containing the Weinberg-Tomozawa term~\cite{WeinbergTomozawa} 
and other $\pi^{2n}NN$ 
terms, $g_A$ is the axial-vector constant, $f_\pi$ the pion decay constant.
Furthermore,
\[
u_\mu = i\left( u^\dagger \partial_\mu u- u \partial_\mu
u^\dagger\right)\quad\mbox{and}\quad
\chi_+ = u^\dagger \chi u^\dagger + u \chi^\dagger u 
\]
are the chiral vielbein and the mass term, respectively,
with $\chi=2B{\mathcal M}$, where ${\mathcal M}$ is the quark mass matrix
and $B$  is proportional to the $SU(2)$ quark condensate 
in chiral limit.
In the isospin symmetric case one may write
$
\chi = m_\pi^2 \, \mathbf{1}
$.

In order to investigate the dependence
of our results
on the choice made for the pion field $\pi$ (= $\vec\tau\cdot\vec\pi$ in 
terms of the Pauli matrices  for isospin),
we start from the following general expression
for the chiral matrix $U=u^2$~(see Appendix~\ref{app:chiral_matrix})
\begin{equation}
 U = \exp \left(\frac{i}{f_\pi}(\vec \tau \cdot \vec \pi) g(\pi ^2/f_\pi^2)
 \right) \ .
\label{Umat}
\end{equation}
Here  $g(\pi^2/f_\pi^2)$ is an arbitrary regular function with 
$g(0)=1$. For
our purposes it is sufficient
to expand $g$ up to second order in the pion field. We may write
\begin{equation}
 g (\pi^2/f_\pi^2) 
 = 1+\left(\alpha+\frac{1}{6}\right)\frac{{\vec \pi}^2}{f_\pi^2}+\cdots \ .
\end{equation}
Obviously, for $\alpha=-1/6$ we work with $U$ in the so-called 
{\em exponential gauge}.
In the $\sigma$--gauge one uses
\begin{equation}
 U=\sqrt{1-\frac{{\vec \pi}^2}{f_\pi^2}}+\frac{i}{f_\pi}\vec \tau \cdot
 \vec\pi
 =1+\frac{i}{f_\pi}\vec\tau \cdot \vec\pi -
 \frac{1}{2f_\pi^2}{\vec\pi}^2-\frac{1}{8f_\pi^4}{\vec\pi}^4-\cdots \ .
\end{equation}
By explicit evaluation one finds, that $\alpha=0$ reproduces this expression
up to and including terms of order $(\pi / f_\pi)^4$. This is sufficient
for our purposes.  For more
details including a justification of the notion ``gauge choice'' for
the pion parameterizations  
see Appendix~\ref{app:chiral_matrix}.

All building blocks of the chiral Lagrangian may now
be expressed in terms of the field $u$ defined above. One
finds for the operators relevant in this work
\begin{equation}
u_\mu =
-\frac1{f_\pi}\vec \tau \cdot \partial_\mu \vec \pi
-\frac1{2f_\pi^3}\left(2\alpha{\vec \pi}^2 \left(\vec \tau \cdot
\partial_\mu \vec \pi\right)+\left(1+4\alpha\right)
\left(\vec \pi\cdot \partial_\mu \vec \pi\right)\left(\vec \tau\cdot
\vec \pi\right)\right)
+ \cdots
\end{equation}
for the chiral vielbein and
\begin{equation}
\chi_+=m_\pi^2\left(u^\dagger u^\dagger +
uu\right)=m_\pi^2\left(2-\frac{{\vec \pi}^2}{f_\pi^2}-
\frac{{\vec \pi}^4}{4f_\pi^4}\left(1+8\alpha\right) - \cdots \right)
\end{equation}
for the mass term. In both cases terms of
higher order in the pion field were not displayed, since
they are not relevant for the present work.

We use the framework of  heavy-baryon
chiral perturbation theory (HBChPT) \cite{JenMano,BKKM}. It is then
straightforward to find
the Feynman rules
for
the relevant building blocks of the diagrams (see Fig.~\ref{blocks}, 
in all cases the
pion momenta $q_i=(q^0_i,{\vec q}_i)$, 
$i=1,2,3,4$ are chosen as outgoing):

\begin{figure}[t]
\begin{center}
\epsfig{file=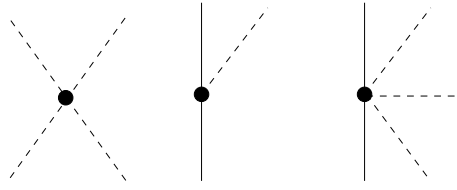,width=6cm}
\end{center}
\caption{The different vertices relevant for this study. Dashed lines
denote pions and solid lines denote nucleons.}
\label{blocks}
\end{figure}

\begin{eqnarray}
\nonumber
i V_{4\pi} &=& \frac{i}{f_\pi^2}\Bigl\{\Bigl[
(q_1+q_2)^2-m_\pi^2+2\alpha\sum_{i=1}^4 
(q_i^2-m_\pi^2)\Bigr]\delta^{ab}\delta^{cd}
 \\
& & \qquad + 
   \Bigl[ \mystack{a b;c d}{12;34} \rightarrow 
          \mystack{ac;bd}{13;24}\Bigr] 
 + \Bigl[ \mystack{a b;c d}{12;34} \rightarrow 
          \mystack{ad;cb}{14;32}\Bigr] 
\Bigr\}\ ,
\label{V4pi} \\
i V_{NN\pi} &=& -\frac{g_A}{2f_\pi}\tau^a (\vec \sigma\cdot {\vec q}_i) \ ,
\label{VNNpi}\\
i V_{NN3\pi} &=& -\frac{g_A}{4f_\pi^3}\left\{\delta^{ab}\tau^c \vec
\sigma\cdot \left[\vec q_1+\vec q_2
+4\alpha(\vec q_1+\vec q_2+\vec q_3)\right]+ \mbox{cyclic}\right\} \ ,
\label{VNN3pi}
\end{eqnarray}
where $\vec\sigma$ is the Pauli-matrix vector for spin and 
$a,b,c,d\in\{1,2,3\}$ are the isospin 
indices.\footnote{In the sigma gauge all vertices 
can be found in appendix A of Ref.~\cite{ulfbible}.}
In addition we need both the pion propagator and the nucleon
propagator. The former is given by the standard expression
$i D_\pi(q)^{ab}= i\delta^{ab}(q^2-m_\pi^2+i\epsilon)^{-1}$.
To leading order
we use for
the latter
\begin{equation}
i S_N(p-q)=\frac{i}{-q_0+i\epsilon} \ .
\label{nprop1}
\end{equation}
We chose the momenta such that the
initial, on-shell, nucleon with momentum
$P^\mu=Mv^\nu+p^\mu$
is pushed off its mass shell by the
emission of a virtual pion with momentum $q$, 
where $v^\mu$ is a four-vector with the properties
$v^2=1$ and  $v^0\geq1$. The standard choice of HBChPT, also used here, 
is  $v^\mu=(1,0,0,0)$. Throughout the paper we follow 
the convention that uppercase nucleon momenta
contain $Mv^\mu$, whereas this term is subtracted out from 
their lowercase counterparts. Note that for
loop momenta temporal ($q_0$) as well as spacial ($q_i$)
components are assumed of order $p_{\rm thr}$, if not stated otherwise, 
as outlined in the introduction.
The residual energy 
$p_0$
of the incoming {\em on-shell} proton, however, is of order 
$\chi_{\rm prod}^2 M\sim m_\pi$ because of the on-shell condition.
Our rule  for the nucleon propagator
is different to the one applied in Refs.~\cite{loop_dmit,fred},
where $i(p_0-q_0+i\epsilon)^{-1}$ is used for the propagator.
It
is justified in the next section
and in Appendix~\ref{app:propagator}.
For a very explicit derivation of the
rules of the heavy baryon formalism we refer to Ref.~\cite{scherer1}
--- see chapter 5.5.6 and  Eq.\,(5.112) 
for another justification that, to leading order in
the $1/M$ expansion, 
$v\cdot p$  has to vanish.

With the building blocks at hand we can now evaluate diagram $(a)$
of Fig.~\ref{loops}. Especially let us focus on those
terms that are proportional to $\alpha$. These read
\begin{eqnarray}
\nonumber
i\tilde
A_{(a)}^{\rm
NLO}&=&-2\frac{i\alpha}{f_\pi^2}\left(\frac{g_A}{2f_\pi}\right)^3\frac{i^4}
{k^2-m_\pi^2}(\vec\sigma_2\cdot \vec k)
\tau_2^c \\
& & \ \times
\int\frac{d^4l}{(2\pi)^4}\frac{\vec \sigma_1\cdot ({\vec p}'-\vec l-\vec p)
\tau_1^a(\vec \sigma_1\cdot \vec
l)\tau_1^b}{(l^2-m_\pi^2)((p'-l-p)^2-m_\pi^2)(l_0+i\epsilon)}i\tilde
V_{4\pi}^{ab cd} \ ,
\end{eqnarray}
where
\begin{equation}
i\tilde
V_{4\pi}^{ab cd}
=(\delta^{ab}\delta^{cd}+\delta^{ac}\delta^{bd}+\delta^{ad}\delta^{bc})
\left[(l^2-m_\pi^2)+((p'\mbox{$-$}l\mbox{$-$}p)^2-m_\pi^2)+(k^2-m_\pi^2)\right] 
 .
\label{v4pi}
\end{equation}
Here  the indices $j$ = 1, 2 of 
the Pauli matrices $\tau_j$ and 
$\sigma_j$  refer to the  left and  right nucleon lines in Fig.~\ref{loops},
respectively;
$l\sim p_{\rm thr}$
denotes the momentum of the pion loop, $p$ and $p'$ are
the momenta
of the incoming and outgoing leg
of the left nucleon line ($j$=1), respectively, whereas
$k$ is the difference of the incoming momentum  
minus the outgoing one of the right nucleon line ($j$=2). 
Note that the temporal components $p_0$, $p_0'$ and  $k_0$ 
of the nucleon momenta or, respectively, nucleon-momentum difference 
scale all as 
$\chi_{\rm prod}^2 M~\sim m_\pi$, whereas their spatial counterparts  
$p_i$, $p_i'$ and  $k_i$
scale as
$p_{\rm thr}$.
The uncontracted index $d$ refers to the isospin of the
produced pion. Its momentum is equal to $k+p-p'$ and is of course on-shell
and scales as $m_\pi$.
The corresponding term for diagram $(b)$ gives
\begin{eqnarray}
\nonumber
i\tilde
A_{(b)}^{\rm NLO}&=&-2i^3\frac{\alpha}
{f_\pi^3}\left(\frac{g_A}{2f_\pi}\right)^3(\vec\sigma_2\cdot \vec k)\,
(\delta^{ab}\tau_2^d+\delta^{ad}\tau_2^b+\delta^{bd}\tau_2^a)
\\
& & \qquad \times
\int\frac{d^4l}{(2\pi)^4}\frac{\vec \sigma_1\cdot ({\vec p}'-\vec l-\vec p)
\tau_1^a(\vec \sigma_1\cdot \vec l)
\tau_1^b}{(l^2-m_\pi^2)((p'-l-p)^2-m_\pi^2)(l_0+i\epsilon)} \ .
\end{eqnarray}
Note that the particular combination of momenta as
it appears in the $\alpha$-dependent
terms of the three-pion vertex is independent of the integration variable
$l$ and was therefore  pulled out of the integral. As a consequence
the integral $\tilde A_{(b)}^{\rm NLO}$ exactly
cancels that part of $\tilde A_{(a)}^{\rm NLO}$ that
corresponds to the last term of Eq. (\ref{v4pi}). What remains to
be studied are the other two terms. Each of them cancels one of
the pion propagators inside the integral.
We get
\begin{eqnarray}
\nonumber
&&i\left(\tilde A_{(b)}^{\rm NLO}+\tilde A_{(a)}^{\rm NLO}\right)
 =-10\frac{i\alpha}{f_\pi^2}\left(\frac{g_A}{2f_\pi}\right)^3
\frac{i^4}{k^2-m_\pi^2}(\vec \sigma_2\cdot \vec k)\tau_2^d \hfill \\
& &
{\times}
\int\frac{d^4l}{(2\pi)^4}\frac{\vec \sigma_1\cdot
({\vec p}'\mbox{$-$}\vec l\mbox{$-$}\vec p)
(\vec \sigma_1\cdot \vec l)}{(l_0+i\epsilon)}
\left\{\frac1{l^2-m_\pi^2}+\frac1{(p'\mbox{$-$}l\mbox{$-$}p)^2-m_\pi^2}\right\}
\ .
\label{firststep}
\end{eqnarray}
Using the variable transformation $l\to l'=p'\!-\!l\!-\!p$ 
in the second term we
find
\begin{eqnarray}
\nonumber
i\left(\tilde A_{(b)}^{\rm NLO}+\tilde A_{(a)}^{\rm NLO}\right)
&=&-10\frac{i\alpha}{f_\pi^2}\left(\frac{g_A}{2f_\pi}\right)^3
\frac{i^4}{k^2-m_\pi^2}(\vec \sigma_2\cdot \vec k)\tau_2^d \hfill \\
& &
\!\!\!\!\!\!\!\!\!\!\!\!\!\!\!\!\!\!\!\!\!\!\!\!\!\!\!\!\!\!\!\!\!\!\!\!\!\!\!
{\times}
\int\frac{d^4l}{(2\pi)^4}\frac{\vec \sigma_1\cdot (\vec w-\vec l)
(\vec \sigma_1\cdot \vec l)}{l^2-m_\pi^2}
\left\{\frac1{l_0+i\epsilon}+\frac1{w_0-l_0+i\epsilon}\right\}
\ ,
\label{intermediate}
\end{eqnarray}
where we defined $w\equiv p'-p$ and renamed $l'$ back to $l$.
The integrand now does not contain the large scale $|\vec p|$
anymore in the denominator. Consequently, $l$ 
will now  be of order $m_\pi$
and no longer of order $p_{\rm thr}$. This is why $w_0$, 
which is also
of order $m_\pi$ (while $|\vec w|\sim p_{\rm thr}$), 
is to be kept in the denominator of the last term.
The angular integration leads to
$$
\vec \sigma_1\cdot ({\vec p}'-\vec l-\vec p)
(\vec \sigma_1\cdot \vec l) \ \to \ 
- {\vec l}^2 = (l^2-m_\pi^2)-(l_0^2-m_\pi^2) \ .
$$
Inserting the first bracket into the above integral leads
to a vanishing result, since the spatial integration
is free of scales. We may therefore write
\begin{eqnarray}
\nonumber
i\left(\tilde A_{(b)}^{\rm NLO}+\tilde A_{(a)}^{\rm NLO}\right)
&=&10\frac{i\alpha}{f_\pi^2}\left(\frac{g_A}{2f_\pi}\right)^3
\frac{i^4}{k^2-m_\pi^2}(\vec \sigma_2\cdot \vec k)\tau_2^d \hfill \\
& &
{\times} w_0
\int\frac{d^4l}{(2\pi)^4}\frac{l_0^2-m_\pi^2}{l^2-m_\pi^2}
\left\{\frac{1}{(l_0+i\epsilon)(w_0-l_0+i\epsilon)}\right\}
\ .
\label{nloadep}
\end{eqnarray}
The only external scales in the integral are $m_\pi$
and $w_0\sim m_\pi$.
In addition,
the integral is quadratically divergent. Therefore, when being
evaluated in dimensional regularization, the resulting expression 
will scale as 
$|\vec k|^{-1}
\times w_0\times m_\pi^2\sim 1/p_{\rm thr}\times m_\pi^3$.

As was explained in Ref.~\cite{mitnorbert}, the leading loops
(including pion-field independent terms) can be well
estimated by identifying all momentum/energy scales by $p_{\rm thr}$.
Thus the $\alpha$-dependent terms of the sum of diagram $(a)$ and
$(b)$ are suppressed by a power of $(m_\pi/p_{\rm
thr})^3=\chi_{\rm prod}^3$ compared to the leading loops that
start to contribute at order NLO.  This implies that the pion-field
dependent terms start to contribute only at order
N$^4$LO.\footnote{The power counting can only account
for a parametric suppression of diagrams relative to each
other. Obviously the expression of Eq.~(\ref{nloadep}) can be enhanced
artificially by choosing a gauge that corresponds to a very large
value $\alpha$. Note that for all standard choices
$|\alpha|\le 1/4$ (see
Eqs. (\ref{exppara})--(\ref{weinpara})). For practical
purposes, the $\sigma$-gauge is of course the most efficient one,
since each $\alpha$-dependent term trivially vanishes
individually.} At this order the sum of Eq.~(\ref{nloadep}) cancels
against the sum of the $\alpha$-dependent contributions of diagrams
$(c)$ and $(d)$ of Fig.~\ref{loops} which are separately of order
N$^4$LO, when the same Feynman rules are applied in the calculation.
This, however, does not exclude the possibility that there may exist
other $\alpha$-dependent terms of order N$^4$LO that result from
subleading Feynman rules.  This we will investigate in the next
section.

\section{Beyond leading order}
\label{subleading}

\begin{figure}[t]
\begin{center}
\epsfig{file=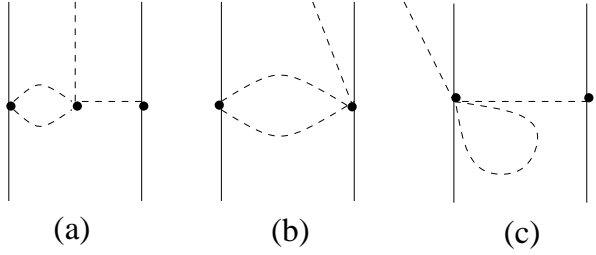,width=8cm}
\end{center}
\caption{Some one-loop diagrams that start to contribute to
to $NN\to NN\pi$ at NNLO, $(a)$ \& $(b)$, and N$^4$LO, $(c)$.
Dashed lines
denote pions and solid lines denote nucleons. The exchange diagrams are not
shown. Note that the diagrams $(b)$ and  $(c)$ result from diagram
$(a)$ under the removal of one internal pion line.
The $\pi\pi NN$ vertex my either be taken from ${\mathcal L}_{\pi
N}^{(1)}$ ---
Weinberg--Tomozawa term~\cite{WeinbergTomozawa} --- or from
those vertices of ${\mathcal L}_{\pi N}^{(2)}$
that depend on the pion momenta ({e.g.} the $c_3$ term) \cite{GSS}.}
\label{loops2}
\end{figure}

We found that to NLO the sum of diagram $(a)$ and $(b)$
(and -- trivially -- the sum of diagram $(c)$ and $(d)$) of Fig.~\ref{loops}
is invariant under the choice of the pion field.
All terms that depend on the pion field vanish to this order.
In this section we investigate the pion-field dependent terms
of the diagrams shown in Fig.~\ref{loops} to NNLO. Please note that
there are several additional diagrams contributing to this order that
are potentially pion-field dependent --- one example being
the so--called football diagrams shown in Fig.~\ref{loops2}.
The proof that to order lower than N$^4$LO, {i.e.} to 
${\mathcal O}(\chi_{\rm prod}^n)$ with $n\leq 4$,
there are no $\alpha$-dependent terms resulting from the additional diagrams
is analogous to the one given
here for the diagrams of Fig.~\ref{loops}
and thus we do not present it in detail.

To order NNLO the only thing that needs to
be considered is the subleading contribution
to the nucleon propagator, which is suppressed by one power in
$\chi_{\rm prod}$
compared to (\ref{nprop1}).\footnote{Of course the subleading
contribution to the propagator can be interpreted  as 
an ${\mathcal O}(q^2/M)$ insertion 
between two  leading-order propagators $(-q_0+i\epsilon)^{-1}$.}
The subleading $\pi NN$ vertices
on the other hand are already
down by $m_\pi/M\sim \chi_{\rm prod}^2$.
There are two pieces to the subleading propagator: one 
from treating $p_0$ as subleading in Eq.~(\ref{nprop1})
\begin{equation}
\nonumber
i\Delta_1
S_N(p-q)\equiv\frac{i}{p_0-q_0+i\epsilon}-\frac{i}{-q_0+i\epsilon}
=-i\frac{p_0}{(q_0-i\epsilon)^2}
\left[1+{\mathcal O}\left({\frac{p_0}{q_0}}\right)\right] \ ,
\end{equation}
and one coming
from the $1/M$ corrections \cite{ulfbible} given by
\begin{equation}
\nonumber
i\Delta_2
S_N(p-q)=\frac{i}{2M}\left(1-\frac{(p-q)^2}{(p_0-q_0+i\epsilon)^2}\right)
=i\frac{(\vec p-\vec q)^2}{2M(q_0-i\epsilon)^2}
\left[1+{\mathcal O}\left({\frac{p_0}{q_0}}\right)\right]
\ .
\end{equation}
Putting both pieces together we get the
following next-to-leading contribution of the nucleon propagator in HBChPT,
using the on-shell condition
$p_0={\vec p}^2/2M+\mathcal{O}({\vec p}^4/M^3)$ and neglecting higher-order 
terms:
\begin{eqnarray}
i\Delta S_N(p-q)=i\Delta_1 S_N(p-q)+i\Delta_2 S_N(p-q)
=i\frac{{\vec q}^2-2\vec p\cdot \vec q}{2M(q_0-i\epsilon)^2} \ .
\label{subleadingprop}
\end{eqnarray}
This illustrates nicely why $p_0$ should be treated as
order $\chi_{\rm prod}^2M$: if the  nucleon leg attached
to the propagator is on-shell,
the $p_0$ term gets canceled by the ${\vec p}^2/(2 M)$ term
of the $1/M$ corrections, as soon as both contributions
are treated on equal footing.
Note 
that each of the two steps of the derivation was based on  
$p_0/q_0\sim \chi_{\rm prod}$, whereas the total 
result holds in general.
Therefore we present
in Appendix~\ref{app:propagator}  a straight forward
derivation of this result --  based on the covariant propagator -- 
that still is valid even when  
$q_0\sim |\vec q|\sim p_0 \sim m_\pi$.

Using Eq.~(\ref{subleadingprop}) for the nucleon propagator
we get for the NNLO contribution of the $\alpha$-dependent terms
of diagram $(a)$ of Fig.~\ref{loops} with $\tilde V_{4\pi}^{ab cd}$ 
as in Eq.~(\ref{v4pi}):
\begin{eqnarray}
\nonumber
i\tilde
A_{(a)}^{\rm NNLO}&=&
-2\frac{i\alpha}{f_\pi^2}\left(\frac{g_A}{2f_\pi}\right)^3\frac{i^4}
{k^2-m_\pi^2}(\vec\sigma_2\cdot \vec k)
\tau_2^c\int\frac{d^4l}{(2\pi)^4}
\left(\frac{2\vec l\cdot \vec p+{\vec l}^2}{2M}\right) \\
& & \qquad  \times
\frac{\vec \sigma_1\cdot ({\vec p}'-\vec l-\vec p)
\tau_1^a(\vec \sigma_1\cdot \vec l)\tau_1^b}
{(l^2-m_\pi^2)((p'-l-p)^2-m_\pi^2)(l_0+i\epsilon)^2}i\tilde
V_{4\pi}^{ab cd} \ . 
\label{tildeannlo}
\end{eqnarray}
As before, the integral
that emerges when introducing the last term of  Eq.~(\ref{v4pi}) into
Eq.~(\ref{tildeannlo}) gets canceled by the corresponding term
for diagram $(b)$ and we refrain from showing the expression explicitly.
After the same variable transformation
($l\to l'=p'-l-p$)  as above, the remainder reads
\begin{eqnarray}
\nonumber
&& i\left(\tilde A_{(b)}^{\rm NNLO}+\tilde A_{(a)}^{\rm NNLO}\right)
=-10\frac{i\alpha}{f_\pi^2}\left(\frac{g_A}{2f_\pi}\right)^3
\frac{i^4}{k^2-m_\pi^2}(\vec \sigma_2\cdot \vec k)\tau_2^d
\frac1{2M}
  \\
& &
\times \int\frac{d^4l}{(2\pi)^4}
\frac{\vec \sigma_1\cdot (\vec w\mbox{$-$}\vec l)
(\vec \sigma_1\cdot \vec l)}
{l^2-m_\pi^2}\left\{\frac{2\vec l\cdot \vec p+{\vec l}^2}
{(l_0+i\epsilon)^2}
+\frac{2(\vec w\mbox{$-$}\vec l)\cdot \vec p+
(\vec w\mbox{$-$}\vec l)^2}{(w_0-l_0+i\epsilon)^2}\right\}
,
\label{tildeapbnnlo}
\end{eqnarray}
again using $w=p'- p$. As before,
this integral diverges (at least) quadratically
with the only scale in the denominator given by  $m_\pi\sim w_0$. In
addition
there is now an overall scale of order ${\vec p}^2/M$ present, which
is also of order $m_\pi$. Therefore the integral given
in Eq.~(\ref{tildeapbnnlo}) also starts to contribute only at order N$^4$LO.
Again this sum cancels against the summed $\alpha$-dependent
contributions of diagram $(c)$ and $(d)$
of Fig.~\ref{loops}, if
the subleading nucleon propagator (\ref{subleadingprop}) is inserted
into the latter diagrams.

Finally note that the next-to-subleading
correction to the nucleon propagator and
vertices would necessarily involve an additional factor
$\chi_{\rm  prod}$ relative to the above presented N$^4$LO result.
The summed
$\alpha$-dependent contributions of diagram $(a)$ and $(b)$ (and of 
$(c)$
and $(d)$)
of Fig.~\ref{loops}
resulting from this next-to-subleading order should therefore contribute
only at order N$^5$LO. In other words, the proof that the $\alpha$-dependent
terms in the sum of all diagrams of  Fig.~\ref{loops} cancels to order
N$^4$LO is now complete.

\section{Conclusions}
\label{conclusion}

Of course, the fact that there is a cancellation of the summed
$\alpha$-dependent terms of the four diagrams of Fig.~\ref{loops} does
not come as a surprise, see Ref.~\cite{novel}, since these terms
would cancel also in a relativistic 
calculation of the
type \cite{GSS} where the nucleon propagator (\ref{nprop1}) is
replaced by the non-expanded covariant form (\ref{covprop}) and where
the terms $-\vec \sigma \cdot {\vec q}_i$ appearing in the vertices
(\ref{VNNpi}) and (\ref{VNN3pi}) are replaced by their covariant
Dirac-analogs $\gamma_\mu\gamma_5 (q_i)^\mu$. In fact, this
cancellation between the $\alpha$-dependent contributions of diagram
$(a)$ on the one hand and the ones of diagrams $(b)$, $(c)$ and $(d)$
on the other hand is solely based on the cancellations of the inverse
pion-propagators appearing in Eq.~(\ref{v4pi}) and the various pion
propagators appearing in diagram $(a)$ which all are of covariant
nature -- even in HBChPT.  The point, however, is that now it is clear
that this cancellation is also consistent with the two-scale expansion
scheme of Refs.\,\cite{pwaves,mitnorbert}: (i) we have explicitly
shown that the pion-field dependent contributions of the diagrams
$(a)$ and $(b)$ of Fig.\,\ref{loops}, although both are NLO diagrams,
cancel at NLO and at N$^2$LO when calculated with leading and
next-to-leading input, respectively, for the nucleon-propagators and
vertices. (ii) We have shown that the remainders are only of N$^4$LO,
the very same order at which the diagrams $(c)$ and $(d)$ start to
contribute. (iii) At this order, the pion-field dependent
contributions of the sum of diagram $(a)$ and $(b)$ indeed cancel
against the corresponding contributions of diagram $(c)$ and $(d)$.
(iv) We have argued that further subleading orders of the nucleon
propagator and vertices will lead to pion-field dependent terms that
are at least of N$^5$LO.

These results can be generalized in the following way:
as long as
the order in the  expansion of the
nucleon propagator in diagram $(a)$ matches those of diagrams $(b)$, $(c)$
and $(d)$ and as long as
the order in the expansion of the $NN\pi$ vertex matches those
of the $NN3\pi$ vertex, the following cancellations are bound to happen:
first, the cancellation between the $\alpha$-dependent
contribution of diagram $(b)$ and the one of diagram $(a)$ that results
from the insertion of the last term of Eq.~(\ref{v4pi});
at this stage
the remainder of the $\alpha$-dependent
contribution of diagram  $(a)$ has now the same order in the 
two-scale expansion
as the  $\alpha$-dependent contributions of diagram $(c)$ and $(d)$
calculated with the same input; secondly,
since the cancellation is based on covariant input from the (inverse)
pion-propagators and since the rest of the input is the same,
the sum of these remaining  $\alpha$-dependent contributions has to vanish.
Of course, at the same order in the chiral expansion, say at N$^n$LO with
a fixed $n>4$,
the diagrams of Fig.~\ref{loops} might generate additional $\alpha$-dependent
terms
resulting  from further subleading
orders in the expansion of the nucleon propagators and vertices,
as it was {e.g.} the
case at the leading and subleading order in the expansion of the nucleon
propagator. Nevertheless, for the same reasons as above, also
these additional contributions have to sum to zero. Eventually at an even
higher order  in the expansion of the nucleon propagator and vertices
no more $\alpha$-dependent terms of  N$^n$LO 
can appear in the
summation; instead contributions of the next order $n+1$
will arise which
again sum to zero and so on. As indicated, our proof linked to the
diagrams of Fig.\,\ref{loops} can easily be generalized to other classes of
potentially pion-field dependent diagrams as {e.g.} given 
in Fig.\,\ref{loops2}.  
In summary, the two-scale expansion
scheme of Refs.\,\cite{pwaves,mitnorbert} is consistent with pion-field
independence to all orders in the expansion. 

As by-products of the investigation we could show that
the {\em parameterizations} of the pion field indeed correspond to 
{\em gauge choices}, 
and could clarify the structure of the
heavy-baryon propagator connected to an on-shell
nucleon leg. Contrary to a naive interpretation of the heavy-baryon rules the
on-shell residual energy of the external nucleon is of the same
order as the kinetic recoil term of the nucleon -- in fact, to the very same 
order, 
they  cancel each other.

\section*{Acknowledgements}
We thank Fred Myhrer for fruitful 
discussions at an early stage of this
research.

\appendix
\section{Reparameterizations of the chiral matrix $U$}
\label{app:chiral_matrix}

The theorem that {\em on-shell} matrix elements do not dependent on
the specific parameterization of the local interpolating field(s) has
a long history reaching back to the LSZ reduction formula~\cite{LSZ55} and the 
work of Haag~\cite{haag58}, see also 
Refs.~\cite{borchers60,chisholm61,kamefuchi61,ruelle62} etc.  
In the context of non-linear
realizations of chiral Lagrangians this general theorem of axiomatic
field theory was confirmed in Ref.\,\cite{CWZ69}. The more
restricted question of the general reparameterizations of the chiral
matrix $U$ for the chiral group $SU(2)\times SU(2)$, more specifically,
the general repara\-meteri\-zation of the pion field under nonlinear
transformations induced by chiral $SU(2)\times SU(2)$ was first
studied by Weinberg~\cite{wein68}. {}From the parity of the pion and the
transformation properties of the pion field under vector
and axial--vector
transformations
combined with Jacobi-identity constraints
Weinberg could show that the
most general redefinition of the nonlinearly realized pion field is of
the form
\begin{equation}
{\pi'}^a = \pi^a g(\pi^2), 
\quad 
a\in\{1,2,3\}
\end{equation}
where the $\pi^a$'s are the usual isospin components of the pion
field $\pi\equiv\vec\tau\cdot\vec\pi= \sum_{a=1}^3\tau^a \pi^a$
and where $g(\pi^2)$ is regular 
in $\pi^2={\vec \pi}^2=\sum_{a=1}^3
(\pi^a)^2$.
In terms of the chiral matrix and the dimensionful version of the 
pion 
field  this  corresponds to
\begin{equation}
U' \equiv \exp\left (\frac{i}{f_\pi} \vec\tau \cdot  {\vec{\pi'}}
\right)=
\exp\left (\frac{i}{f_\pi} \vec\tau \cdot \vec\pi\,
g(\pi^2/f_\pi^2)\right) \ .
\label{Umat1}
\end{equation}
This is the result of Eq.\,(\ref{Umat}) under the additional condition that
$g(0)=1$, which follows from fixing the wave function normalization
of the pion at tree-level or, in other words, the
free-particle part of the Lagrangian (\ref{LGSS}), 
see Refs.~\cite{chisholm61,kamefuchi61}.

The various known parameterizations, the exponential one, the so-called
$\sigma$-gauge, the Weinberg one~\cite{wein68},
etc.\ follow from (\ref{Umat1}) with the
help of the following choices of $g(\pi^2/f_\pi^2)$ functions:
\begin{eqnarray}
    g(\pi^2/f_\pi^2)&=& 1 \qquad\qquad\qquad\qquad\quad 
     (\mbox{exponential parameterization}),
\label{exppara}\\
    g(\pi^2/f_\pi^2) &=& 
\frac{1}{\sqrt{\pi^2/f_\pi^2}}
\arcsin \left(\sqrt{\pi^2/f_\pi^2}\,\right)
=  1 +\frac{\pi^2}{6f_\pi^2} 
+ \cdots\ (\sigma\mbox{-gauge}),\label{G-sigma} 
\\
    g(\pi^2/f_\pi^2) &=& \frac{1}{\sqrt{\pi^2/f_\pi^2}}
      \arcsin\left(\frac{\sqrt{\pi^2/f_\pi^2}}
      {1+\pi^2/(4 f_\pi^2)}\right)
       = 1 -\frac{\pi^2}{12 f_\pi^2} 
+ \cdots\nonumber\\
&&
\qquad\qquad\qquad\qquad\qquad\ \ (\mbox{Weinberg parameterization}).
\label{weinpara}
\end{eqnarray}
In fact,
the transformation (\ref{Umat1}) can be simplified by the following
rescaling
\begin{equation}
U' = \exp\left (\frac{i}{f_\pi}  \vec\tau \cdot \vec\pi\,
g(\pi^2/f_\pi^2)\right)
=\exp\biggl( {i} \vec\tau \cdot  \hat{\vec\pi}\, 
 F\Bigl(\sqrt{\pi^2/f_\pi^2}\,\Bigr)
\biggr)
\label{Umat2}
\end{equation}
where $\hat{\vec \pi}={\vec \pi}/\sqrt{\pi^2}$
is the pion unit vector  in isospin-space and
$F(x)=x g(x^2)$ is an odd analytic function of the variable $x$ with a
normalized
first coefficient in the
Taylor expansion, $F(x) = x + \sum_{n=2}^\infty c_{2n-1}x^{2n-1}$, see
{e.g.} \cite{delorme96,chanfray96}.
In terms of this function
the various parameterizations become especially simple
\cite{delorme96,chanfray96}:\footnote{In 
$SU(2)$:
$\exp\Bigl(i \vec\tau\cdot \hat{\vec\pi} 
     F\bigl(\!\sqrt{\pi^2/f_\pi^2}\,\bigr)\Bigr)\! =\! 
\cos\Bigl(\!F\bigl(\!\sqrt{\pi^2/f_\pi^2}\,\bigr)\Bigr) 
+ i \vec\tau \cdot\hat{\vec\pi}
  \sin\Bigl(\!F\bigl(\!\sqrt{\pi^2/f_\pi^2}\,\bigr)\Bigr)
$. \label{footsu2}}
\begin{eqnarray}
  F(x) &=& x \qquad\qquad\qquad\qquad\qquad\quad    
  (\mbox{exponential parameterization}),\\
  F(x) &=& \arcsin(x)\qquad\qquad\qquad\quad\;\    
 (\sigma\mbox{-gauge parameterization}),
\label{F-sigma}\\
  F(x) &=& \arcsin\left(x/(1+x^2/4)\right)\qquad  
(\mbox{Weinberg parameterization}).
\end{eqnarray}
In fact, with the help of the machinery of  Ref.\,\cite{CWZ69}
the various  parameterizations can be transformed into each other
by the following axial  gauge transformations
\begin{equation}
U \to U' = U_A(\pi) \,U \,U_A(\pi) \label{Ugauge}
\end{equation}
in terms of the local $SU(2)$ matrix
$
     U_A(\pi) =\exp\left((i/{2f_\pi})  \vec\tau \cdot \vec\pi\, \left(
g(\pi^2/f_\pi^2)-1\right) \right)\ . 
$
The backtransformation
follows
then from the inverse gauge transformation
$U' \to U = U_A(\pi)^\dagger \,U' \,U_A(\pi)^\dagger$.
Transitions between other representations or
{\em gauges}
can be found as
compositions of gauge transformations from and to the exponential gauge,
say. The $\sigma$-{\em gauge} indeed results from a gauge
transformation (\ref{Ugauge}) of the exponential ``gauge''
$U=\exp\left(i \vec\tau\cdot\vec \pi/f_\pi\right)$
when the {\em gauge choice} (\ref{G-sigma}) is inserted into 
$U_A(\pi)$.
In addition, the various
parameterizations of the matrix $u=\sqrt{U}$ transform
into each other as
\begin{equation}
  u \to u' =  U_A(\pi) \,u \,h(U_A,\pi)^{-1}
           = h(U_A,\pi) \,u \, U_A(\pi), \label{ugauge}
\end{equation}
where  $h(U_A,\pi)\in SU(2)_{V}$  is the so-called
``compensator'' or ``hidden'' matrix \cite{CWZ69}
which cancels in 
$U'$ = $u'u'$ = $U_A(\pi) u u  U_A(\pi)$ and in the Lagrangian (\ref{LGSS}).

Whereas the transformations of the type (\ref{Umat1}) or (\ref{Umat2}) are
$SU(2)$ specific, 
the gauge transformation as such -- whether in the form (\ref{Ugauge}) or
(\ref{ugauge}) --
can be generalized to $SU(3)$ with a suitably selected $SU(3)$ gauge matrix
$U_A(\pi)\sim 1 + i \frac{\alpha}{2}\pi^3/f_\pi^3 +\cdots$
that does not spoil the wave function
normalization  at tree level, where
$\pi = \sum_{a=1}^8\lambda^a\pi^a$ in terms of the
Gell-Mann matrices $\lambda^a$.

\section{On the $1/M$ expansion
of the nucleon propagator}
\label{app:propagator}

\begin{figure}[t]
\begin{center}
\psfrag{p}{$P^\mu$}
\psfrag{q}{$q^\mu$}
\psfrag{l}{$P^\mu-q^\mu$}
\epsfig{file=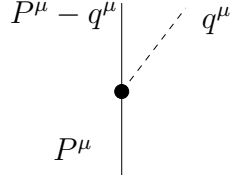,width=3cm}
\end{center}
\caption{Definition of the various momenta
used in Appendix~\ref{app:propagator}. The initial nucleon is supposed
to be on-shell ($P^2=M^2$).}
\label{momdef}
\end{figure}

In the main section the rules of HBChPT were
applied directly. For illustration, we show in this appendix
that the same expressions can be recovered by a
straight forward expansion of the nucleon propagator.
We start from the covariant expression for the nucleon propagator
\begin{equation}
iS_N^{\rm cov}(P-q)=i\frac{P\!\!\!/-q\!\!\!/+M}{(P-q)^2-M^2 +i\epsilon} \ ,
\label{covprop}
\end{equation}
where the momenta are defined in Fig.~\ref{momdef}.
We now want to expand this propagator in powers of $1/M$. 
The easiest way to proceed is via the decomposition
$$
iS_N^{\rm cov}(K)=i\frac{M}{E_K}\sum_s \left\{
\frac{u(\vec K,s)\bar u(\vec K,s)}{K_0-
E_K+i\epsilon}
+\frac{v(-\vec K,s)\bar
v(-\vec K,s)}{K_0+E_K-i\epsilon}\right\} \ ,
$$
where $E_K=\sqrt{M^2+{\vec K}^2}$.
First observe that the second term, corresponding
to the contribution of anti--nucleons, does
not propagate in HBChPT.
It therefore gets absorbed into
local counter terms at the Lagrangian level.
The spinors get part of the vertex functions
and we may therefore focus on the denominator of the first term.
In the kinematics chosen we have
$
K = P-q
$.
To make contact to the expressions of HBChPT, we
write, in accordance with the notation of the
main text,
$ P^\mu = Mv^\mu + p^\mu$ with $v^\mu =(1,0,0,0)$.
Thus
\begin{eqnarray}
P_0-q_0-E_{( P-q)} &=& M+p_0-q_0-\sqrt{M^2+(\vec p-\vec q)^2}
\nonumber
\\
&=&-q_0+\frac{2\vec p\cdot \vec q-{\vec q}^2}{2M}
+{\mathcal O}\left(\frac{(\vec p-\vec q)^4}{M^3}, \frac{{\vec p}^4}{M^3}
\right) \ ,
\label{P0q0E}
\end{eqnarray}
where the on-shell condition
$p_0={\vec p}^2/2M+{\mathcal O}({\vec p}^4/M^3)$ 
was used in the last 
step.
Observe that $p_0$ has disappeared from the expression (\ref{P0q0E}).
However, this
happens only, if $p_0$ is put into the same order as ${\vec p}^2/2M$,
as advocated in the main section.
In the power counting relevant for pion production, the pion energies in
loops are to be counted as order $p_{\rm thr}$ 
as in case (a) and (b) of Fig.~\ref{loops}. Therefore, 
in line with Eqs.~(\ref{nprop1}) and (\ref{subleadingprop}),
the expression for the propagator in HBChPT
is simply
\begin{eqnarray}
\nonumber iS_N(p-q)&=&\frac{i}{-q_0+i\epsilon}
\left(1-\frac{2\vec p\cdot \vec q-{\vec q}^2}{2M(-q_0+i\epsilon)}
+{\mathcal O}\left(\frac{p^2,p\cdot q,q^2}{M^2}\right)\right)
 \\
&=&
\frac{i}{-v\cdot q \!+\!i\epsilon}\left(1+\frac{2 p\cdot q_\perp-q_\perp^2}
{2M(-v\cdot q\!+\!i\epsilon)}
+{\mathcal O}\left(\frac{p^2,p\cdot q,q^2}{M^2}\right)\right)
\ .
\label{sexpand}
\end{eqnarray}
The last relation refers to the general ``velocity'' case   
($v^2=1$ and $v_0\geq 1$) with 
the definition $q_\perp\equiv q-v (v\cdot q)$ and the on-shell condition
$v\cdot p =-p^2/2M$~\cite{scherer1}.

Note that the above equations hold even for more general
kinematics. In all cases of relevance here, the loop momenta are at
least of the order of the pion mass. Thus the components of $q^\mu$
either scale as $p_{\rm thr}$, as used in the previous paragraph, or
as $m_\pi$ --- as in the integral of Eqs.~(\ref{intermediate}) and
(\ref{nloadep}) or in Fig.~\ref{loops} (c) and (d).  
Let us stress that also in the latter case the
expansion of Eq.~(\ref{sexpand}) holds, since the other terms of order
$m_\pi$, namely $p_0$ and $\vec p^2/2M$, canceled and the remaining
recoil terms are suppressed by at least one power of $\chi_{\rm prod}$.

It is also instructive to derive the $1/M$ expansion
of the propagator directly from the covariant expression
of Eq.~(\ref{covprop}).
Using again the on-shell condition for the incoming nucleon, $P^2=M^2$,
we may write
$P_0=M\left(1+{\mathcal O}({\vec p}^2/M^2)\right)$ -- note  that
$\vec P\equiv \vec p$. 
We are interested in the case $|\vec p|\sim p_{\rm thr}$, where
$p_{\rm thr}$ was defined below Eq.~(\ref{chiprod}).
Therefore
${\mathcal O}({\vec p}^2/M^2)$ corresponds to
${\mathcal O}(\chi_{\rm prod}^2)$. 
We thus
identify $-i/q_0$ as the leading term for the propagator
in accordance with Eq.~(\ref{nprop1}). All other terms
that still appear in the denominator are corrections. After
a Taylor expansion to next--to--leading order we get
\begin{eqnarray*}
&&iS_N^{\rm cov}(P-q)=\frac{i}{-q_0+i\epsilon}\biggl\{ {\textstyle\frac12}
\left(\mathbf 1+\gamma_0\right)
\Bigl(1-
{\frac{2\vec p\cdot\vec q-{\vec q}^2}
{2M(-q_0+i\epsilon)}}
\Bigr)\\
& &\quad -\underbrace{\frac1{2M}\vec \gamma\cdot \left(\vec p-\vec
    q\right)}_{\mbox{into vertices}}
\ \ +\underbrace{\frac{q_0}{2M}\,
{\textstyle\frac12}
\left(\mathbf 1-\gamma_0\right)}_{\mbox{effect of anti--nucleon}}
\biggr\}
\left(1
+{\mathcal O}\left(\frac{p^2,p\cdot q,q^2}{M^2}\right)
\right)
\ .
\end{eqnarray*}
As indicated, this expression contains the leading and
next--to--leading piece of the propagator and, in addition,
a piece that can give
momentum dependence to the vertices (this contribution can be mapped
onto the
effect of the spinors in the previous derivation), and, finally,
a contact term
that is the leading term for the effects of the anti--nucleon in the
intermediate state.


\begin{thebibliography}{00}
\bibitem{garmiz}
  H. Garcilazo and T. Mizutani,  $\pi$NN Systems, World Scientific
  (Singapore,  1990).

\bibitem{report}
  C.~Hanhart,
  Phys.\ Rept.\  {\bf 397} (2004) 155.

\bibitem{wein92}
  S.~Weinberg,
  Phys.\ Lett.\  B {\bf 295} (1992) 114.

 
\bibitem{firsts}
  B.~Y.~Park, F.~Myhrer, J.~R.~Morones, T.~Meissner and K.~Kubodera,
  Phys.\ Rev.\  C {\bf 53} (1996) 1519;
  C.~Hanhart, J.~Haidenbauer, M.~Hoffmann, U.-G.~Mei{\ss}ner and J.~Speth,
  Phys.\ Lett.\ B {\bf 424} (1998) 8. 

\bibitem{loop_dmit} 
  V.~Dmitrasinovi\'c, K.~Kubodera, F.~Myhrer and T.~Sato,
  Phys.\ Lett.\ B {\bf 465} (1999) 43.

\bibitem{loop_ando} 
  S.~I.~Ando, T.~S.~Park and D.~P.~Min,
  Phys.\ Lett.\ B {\bf 509} (2001) 253.

\bibitem{bira} 
  T.~D. Cohen, J.~L. Friar, G.~A.~Miller and U. van Kolck, 
  {Phys. Rev.} C {\bf 53} (1996) 2661; 
  C.~da Rocha, G.~Miller and U.~van Kolck,
  Phys.\ Rev.\ C {\bf 61} (2000) 034613. 

\bibitem{pwaves}
  C.~Hanhart, U.~van Kolck, and
  G.~Miller, Phys. Rev. Lett. {\bf 85} (2000) 2905.

\bibitem{mitnorbert}
  C.~Hanhart and N.~Kaiser,
  Phys.\ Rev.\ C {\bf 66} (2002) 054005.

\bibitem{pp2dpi}
V.~Lensky, V.~Baru, J.~Haidenbauer, C.~Hanhart, A.~E.~Kudryavtsev 
and \mbox{U.-G.~Mei{\ss}ner},
  Eur.\ Phys.\ J.\  A {\bf 27} (2006) 37;
 Y.~Kim, T.~Sato, F.~Myhrer and K.~Kubodera,
  arXiv:0704.1342 [nucl-th].

\bibitem{novel}
  V.~Bernard, N.~Kaiser and U.-G.~Mei{\ss}ner,
  Eur.\ Phys.\ J.\ A {\bf 4} (1999) 259.

\bibitem{GSS}
  J.~Gasser, M.~E.~Sainio and A.~\v{S}varc,
  Nucl.\ Phys.\  B {\bf 307} (1988) 779.

\bibitem{WeinbergTomozawa}
  S.~Weinberg,
  Phys.\ Rev.\ Lett.\  \textbf{17} (1966) 616;
  Y.~Tomozawa,
  Nuovo Cim. A \  \textbf{46} (1966) 707.

\bibitem{JenMano}
  E.~Jenkins and A.~V.~Manohar,
  Phys.\ Lett.\  B {\bf 255} (1991) 558.

\bibitem{BKKM}
  V.~Bernard, N.~Kaiser, J.~Kambor and U.-G.~Mei{\ss}ner,
  Nucl.\ Phys.\  B {\bf 388} (1992) 315.

\bibitem{fred}
  F.~Myhrer,
   nucl-th/0611051.

\bibitem{ulfbible}
  V.~Bernard, N.~Kaiser and U.-G.~Mei{\ss}ner,
  Int.\ J.\ Mod.\ Phys.\  E {\bf 4} (1995) 193.

\bibitem{scherer1}
  S.~Scherer,
  Adv.\ Nucl.\ Phys.\  {\bf 27} (2003) 277.


\bibitem{LSZ55}
  H.~Lehmann, K.~Symanzik and W.~Zimmermann,
  Nuovo Cim.\  {\bf 1} (1955) 205; Nuovo Cim. {\bf 6} (1957) 319.


\bibitem{haag58}
  R.~Haag,
  Phys.\ Rev.\ {\bf 112} (1958) 669.

\bibitem{borchers60}
  H.~J.~Borchers,
  Nuovo Cim.\ {\bf 15} (1960) 784.

\bibitem{chisholm61}
  J.~S.~R.~Chisholm, 
  Nucl. Phys. {\bf 26} (1961) 469.

\bibitem{kamefuchi61}
  S.~Kamefuchi, L.~O'Raifeartaigh and A.~Salam,
  Nucl. Phys. {\bf 28} (1961) 529.

\bibitem{ruelle62}
  D.~Ruelle,
  Helv. Phys. Acta {\bf 35} (1962) 147.



\bibitem{CWZ69}
  S.~Coleman, J.~Wess and B.~Zumino,
  Phys.\ Rev.\ {\bf 177} (1969) 2239;
  C.~G.~Callan, S.~Coleman, J.~Wess and B.~Zumino,
  Phys.\ Rev.\ {\bf 177} (1969) 2247.

\bibitem{wein68}
  S.~Weinberg,
  Phys.\ Rev.\  {\bf 166} (1968) 1568.

\bibitem{delorme96}
  J.~Delorme, G.~Chanfray and M.~Ericson,
  Nucl.\ Phys.\ A {\bf 603} (1996) 239.

\bibitem{chanfray96}
  G.~Chanfray, M.~Ericson and J.~Wambach,
  Phys.\ Lett.\ B {\bf 388} (1996) 673.

\end{thebibliography}
\end{document}